# Remotely sensing stress evolution in solids: a passive approach to earthquake monitoring

**Nader Shakibay Senobari[1]***

[1]Department of Computer Science and Engineering, University of California, Riverside; Riverside, CA, 92521, USA.

*Corresponding author. Email: nshak006@ucr.edu

## Abstract

Stress evolution governs material failure across scales—from microscopic fractures to large earthquakes—yet direct observation of its dynamics in natural systems has remained elusive. Laboratory experiments using active ultrasonic measurements have shown that seismic velocity and attenuation are sensitive to stress, but such monitoring has not previously been achievable remotely or passively.

Here, we introduce a new stress-sensitive frequency-domain transform that enables passive monitoring of stress evolution through ambient seismic or acoustic noise. The method quantifies relative energy shifts between adjacent frequency bands, capturing subtle changes in wave-propagation properties linked to evolving shear and normal stress. Applied across scales—from laboratory stick-slip and slow-slip experiments to natural fault systems including the 2018 Kīlauea collapse, Cascadia slow-slip episodes, and major earthquakes such as the 2011 Tōhoku, 2010 Maule, 2002 Denali, and 2023 Turkey–Syria events—the transform consistently reveals distinctive precursory trajectories and stress-cycle patterns.

These results demonstrate, for the first time, that stress evolution within solids can be remotely and passively monitored, bridging laboratory rock physics and large-scale seismology, and offering a new foundation for real-time tracking of fault mechanics and earthquake preparation.

## Introduction

Predicting when and where a large earthquake will occur remains one of geophysics' most formidable challenges. While modern earthquake early warning (EEW) systems can issue alerts seconds before damaging ground motion arrives [1,2], and probabilistic seismic hazard assessments (PSHAs) estimate long-term rupture likelihoods based on historical and geological data [3–5] neither approach addresses the critical gap of short-term earthquake forecasting—spanning hours to days before rupture. This gap has spurred global interest in identifying subtle precursory signals, such as foreshocks and slow slip events (SSEs), which sometimes precede major earthquakes in both time and space [6–10].Yet, these signals are often elusive, inconsistent, or entirely absent, raising the possibility that direct event-based analyses may not be the optimal domain for detecting stress-induced changes in the crust.

An alternative lies in the continuous seismic wavefield itself. Both laboratory and field studies suggest that stress accumulation within a fault zone alters the mechanical properties of surrounding materials, modulating wave velocity, attenuation, and scattering. These stress-related changes affect the spectral and temporal properties of seismic waves—particularly ambient noise—even before macroscopic failure occurs [11–15]. In laboratory experiments, researchers have used active acoustic sources and sensitive sensors to track evolving moduli and damage states leading up to failure [16,17]. However, translating such insights to tectonic-scale systems remains challenging, due to uncontrolled source conditions, signal contamination, and geological heterogeneity.

Passive methods—such as ambient noise interferometry and repeating-earthquake analysis—seek to detect changes in seismic velocity or attenuation without requiring controlled sources [13,14]. While these techniques have provided useful constraints on long-term fault zone evolution, they often lack the temporal resolution to capture short-term dynamics preceding large earthquakes. Moreover, most precursor studies emphasize changes in seismicity or source characteristics, whereas stress-altered wave propagation paths—the so-called "path effects"—remain an underused resource for detecting fault loading in real time.

Recent advances in machine learning have reignited hope for forecasting fault failure. In laboratory-scale analogs, data-driven models have successfully predicted failure timing using only continuous acoustic emissions, by extracting subtle spectral and statistical features imperceptible to human analysts [18–20]. Some recent efforts have extended this strategy to tectonic earthquakes, with mixed results [21,22]. These models often struggle with generalization and interpretability, and rarely illuminate the underlying physical processes that govern failure.

Here, we introduce a new approach for tracking stress evolution in faults using a custom-designed frequency-domain transformation applied to ambient seismic noise. The transformation is crafted to enhance subtle, path-sensitive spectral changes while suppressing variability associated with source intensity, location, and other non-tectonic influences. Drawing on contrast normalization techniques used in image processing, this transform enables high-temporal-resolution tracking of stress-related wavefield changes—without requiring stacking, interferometry, or active sources. It was empirically developed by isolating features that reflect stress evolution in both laboratory rock deformation experiments and real-world seismic data.

We tested this method across a broad range of scales and tectonic settings—from centimeter-scale laboratory experiments to eight major earthquakes, including the 2011 Tōhoku (Mw 9.0), 2010 Maule (Mw 8.8), 2019 Peru (Mw 8.0), 2002 Denali (Mw 7.9), 2023 Turkey–Syria (Mw 7.9), 2015 Gorkha (Mw 7.8), 2018 Kīlauea (Mw 6.9), and 2016 Pawnee (Mw 5.9) events. In every case, the transformation reveals abnormal patterns that potentially indicate precursory spectral signatures—often emerging hours to days before failure—that mirror the stress-loading trends observed in laboratory settings.These results show that ambient seismic noise most-likely contains measurable information about the evolving stress state of faults. By isolating this signal through a simple yet robust transform, we open a new observational window into earthquake nucleation—bringing the longstanding goal of short-term earthquake forecasting within reach.

## Results

### *The novel stress-sensitive transformation*

The goal of the transformation introduced here is to amplify stress-sensitive changes in seismic wavefields while suppressing variability associated with source properties. Seismic waves are shaped by both source characteristics and their propagation path, but conventional signal processing techniques often entangle these influences. To disentangle them, we designed a transformation that responds linearly to changes in source intensity but nonlinearly to changes in the medium.

The transformation compares spectral energy between two adjacent, narrow frequency bands using a logarithmic ratio. For a given time window starting at time $t_0$, and a central frequency $f$ with bandwidth $\delta f$, we define the transformation as:

$$SS_F(t_0, f) = \log\left(\frac{P(t_0, f, \delta f)}{P(t_0, f + \Delta f, \delta f)}\right) \quad (1)$$

Here, *SSF* denotes the Shakibay Senobari Frequency-domain Transform, and $P(t_0, f, \delta f)$ is the average power spectral density in the band centered at $f$, estimated via Welch's method. The denominator represents the neighboring band at $f + \Delta f$. This ratio effectively captures the local logarithmic slope of spectral energy across adjacent bands.

By emphasizing relative changes rather than absolute amplitudes, the transform acts as a nonlinear filter tuned to stress-related wavefield modifications, such as those caused by scattering, attenuation, or microcrack evolution. Crucially, this formulation suppresses broadband fluctuations associated with variable sources or noise bursts, and is robust to station gain or site-specific effects.

The frequency offset, $\Delta f$, and the bandwidth, $\delta f$, are user-defined hyperparameters that control the resolution and sensitivity of the transformation. The parameter $\Delta f$ determines the spacing between adjacent frequency bands and governs the scale at which spectral slope is measured, while also allowing suppression of source-induced nonlinearities through appropriate tuning. The bandwidth $\delta f$ controls the resolution of spectral power estimates and affects the sensitivity to frequency-dependent changes in the medium. Details regarding the specific values of $\Delta f$ and $\delta f$ used in this study—along with full data preprocessing and transformation procedures—are provided in the *Materials and methods* section.

Although the frequency-domain transformation introduced here was developed independently for seismic applications, it is conceptually related to approaches employed in other disciplines. In audio signal processing, spectral slopes are used for automatic detection of emotional stress from speech [23]. In biomedical signal analysis, slope-based spectral features form human brain EEG data are shown to be a reliable indicators of consciousness levels and sleep stages [24,25]. In remote sensing, dual-frequency brightness temperature ratios help infer surface emissivity and atmospheric attenuation [26]. The convergence of these methods across diverse fields underscores the broader utility of logarithmic frequency-ratio transforms for isolating subtle, path-sensitive changes in complex wavefields.

*Fig. 1* demonstrates the output of the SSF transform across a wide frequency range. The resulting time-frequency representations can be aggregated to form a broadband indicator of stress evolution or used to track spectral slope variations in specific frequency bands.

To validate the method, we first apply the transform to controlled laboratory data in which shear and normal stresses are independently measured. These rock deformation experiments offer a critical ground-truth testbed for assessing the sensitivity and physical interpretability of the transform before applying it to tectonic-scale earthquakes.

Below, we first present results from laboratory experiments, followed by nature examples that suggest the potential for establishing a new paradigm and reviving hope for reliable short-term earthquake forecasting.

***Stress evolution signatures in laboratory transforms***

To assess the sensitivity of the new transformation to evolving fault stress, we applied it to acoustic-emission (AE) data from two laboratory experiments (P4581 and P5198) conducted at the Penn State Rock Mechanics Laboratory. These experiments simulate fault slip using different granular materials under controlled stress conditions. In P4581, 100–150 μm glass beads were sheared under normal stresses ranging from 2 to 8 MPa, whereas in P5198, finer quartz powder (~10 μm) was used under higher normal stresses (6–11 MPa). Shear stress and displacement were recorded at 1 kHz, and continuous AE signals were captured at 4 MHz via embedded sensors. Both datasets document the complete stress cycle—from stable sliding to stick–slip failure—providing a high-resolution view of stress accumulation and release.

To explore how the transformation responds under controlled loading, we analyzed time intervals from both experiments that captured multiple cycles of loading and failure under varying normal stress conditions.
Figure 1A shows that the passive transform derived from ultrasonic noise reproduces these stress cycles with striking clarity in P4581, which represents a fast-slip scenario. Low-frequency components evolve nearly linearly with shear stress during loading and drop sharply at failure, whereas higher-frequency components exhibit baseline shifts consistent with normal-stress variations. Together, these frequency-dependent responses enable passive, simultaneous tracking of both shear and normal stress evolution throughout the loading–unloading cycle.

In contrast, experiment P5198 (Fig. 1B) shows a slower and more gradual evolution of the transform, consistent with a slow-slip regime. Across most frequencies, the transform remains sensitive to shear-stress loading and time to failure, but the stacked transforms display an exponential trajectory toward a global minimum prior to slip, followed by a delayed recovery—unlike the abrupt recovery seen in P4581. Similar curvature has been reported in velocity changes during other slow-slip laboratory experiments (refs). In the local stacks (lower panels), low-frequency transforms are most sensitive immediately before and after slip, showing a flat baseline at other times, while higher frequencies show gradual variations through the full cycle. At certain frequencies, such as 378 kHz, the transform captures both shear- and normal-stress effects, with a slight post-slip baseline shift coinciding with the increase in normal stress from 10 MPa to 11 MPa.

### *Stress evolution signatures at tectonic scales*

Figure 2 demonstrates that comparable cyclic stress behavior occurs in natural systems across both fast and slow slip regimes. At tectonic scales, the 2018 Kīlauea caldera collapse sequence (Fig. 2A) reveals that low-frequency components of the transform rise almost linearly between successive collapses and drop abruptly at failure—closely matching the laboratory fast-slip pattern of Fig. 1A. Higher frequencies develop progressive baseline shifts after several collapses, likely reflecting evolving normal stress in the collapsing structure. The recurrence interval between collapses also shortens from roughly two days to about one day, consistent with accelerated stress accumulation.

For the slow-slip case (Fig. 2B), we applied the same analysis to the well-studied Cascadia episodic tremor and slip (ETS) events. The transform evolves smoothly during the inter-ETS phase, then undergoes a sharp, exponentially decaying drop just prior to slow slip, followed by recovery during slip. This pattern closely mirrors the laboratory slow-slip behavior in Fig. 1B. These parallels indicate that stress-controlled modulation of passive wavefields operates universally across scales spanning more than twelve orders of magnitude—from centimeter-scale laboratory faults to kilometer-scale subduction interfaces.

### *Scale-invariant transform patterns*

When comparing laboratory and tectonic fast-slip cases without stacking (Fig. 3), both datasets display nearly identical fine-scale features, including linear and arc-like trajectories, asynchronous extrema across frequencies, and nonlinear temporal evolution. Despite event magnitudes differing by roughly $10^{10}$, the unstacked transforms retain scale-invariant signatures of shear-stress loading and release. In both the laboratory (P4581) and Kīlauea datasets, lower frequencies show a gradual linear decrease toward a minimum immediately before failure, whereas slightly higher frequencies exhibit minima that occur earlier, with increasing time lags relative to failure. This frequency-dependent time lag provides a physical basis for the observed predictive capability of machine-learning models that infer time to failure from acoustic or seismic records.

In natural fault systems, complete loading–failure cycles are rarely observable, as recurrence intervals can span centuries. This poses two challenges: (1) historical seismic archives may not capture an entire stress cycle, and (2) detecting small precursory changes over long durations requires high sensitivity. Nevertheless, the examples presented here demonstrate that passive monitoring of stress evolution is feasible. The goal of this study is not to define a predictive framework for earthquake precursors, but rather to introduce a new passive method for monitoring stress-dependent changes in wave propagation. The preliminary results presented below suggest that such transforms can reveal abnormal or precursory behaviors preceding major earthquakes, motivating systematic future validation and generalization tests.

### *Nature case studies examples*

To test the scalability of the method, we applied the transformation to seismic data from eight well-documented earthquakes across diverse tectonic settings (Figure 4 and 6).

- *2016 Pawnee, Oklahoma:* A magnitude 5.8 earthquake struck on September 3, 2016, marking the largest recorded event in Oklahoma. It occurred within an area of increasing seismicity associated with wastewater injection [27].

- *2023 Turkey–Syria earthquake sequence:* On February 6, 2023, a magnitude 7.8 mainshock was followed by a magnitude 7.5 aftershock, causing widespread damage along the East Anatolian Fault. This sequence produced strong ground motion and foreshock activity across a complex fault system [28].
- *2015 Gorkha, Nepal:* A magnitude 7.8 earthquake ruptured the Main Himalayan Thrust on April 25, 2015. The event caused significant loss of life and exposed the vulnerability of densely populated regions near active continental thrust systems [29].
- *2011 Tōhoku, Japan:* On March 11, 2011, a magnitude 9.0 megathrust event ruptured offshore northeastern Japan. The earthquake generated a massive tsunami and was associated with a well-instrumented rupture along the Pacific–North American plate interface [30].
- *2018 Kīlauea, Hawaiʻi:* On May 4, 2018, a magnitude 6.9 earthquake occurred on the volcano's south flank, coinciding with the onset of a major eruption. More than 60 repetitive Mw ~5 collapse events followed over the subsequent months, linked to rapid magma withdrawal beneath the summit. The event sequence offers an unprecedented dataset linking magmatic, seismic, and structural processes [31].
- *2002 Denali, Alaska:* A magnitude 7.9 strike-slip earthquake ruptured the central Denali Fault on November 3, 2002, producing strong ground motion across Alaska. The rupture began on the Susitna Glacier fault and transitioned onto the main Denali fault, making it an ideal testbed for crustal stress evolution [32].
- *2010 Maule, Chile:* On February 27, 2010, a magnitude 8.8 subduction megathrust earthquake struck offshore central Chile. The event produced broad-scale uplift, triggered tsunamis, and is one of the most comprehensively recorded megathrust events globally [33].
- *2019 Peru:* On May 26, 2019, a magnitude 8.0 intermediate-depth (~120 km) intraslab earthquake struck northern Peru. Despite its large size, the depth limited surface shaking compared to shallow megathrust events. The event provides a rare opportunity to evaluate stress evolution and precursor detectability in deep subduction-zone seismicity[34].

These case studies span a wide spectrum of geologic environments—from induced seismicity and volcanic rifting to continental and subduction faulting. The diversity in faulting styles, stress regimes, and available seismic instrumentation offers a rigorous test for the generalizability of the proposed transformation across scale and setting.

***Potential Precursory signatures in natural events***

While prior laboratory studies have reported velocity changes using active sources [16,35,36] and some machine learning approaches have shown promising performance [18,21], transferring these results to tectonic-scale settings has remained challenging due to the complexity of real seismic environments and the lack of active control. Here, we demonstrate for the first time that the arc-shaped and acceleration-toward-extrema features observed in laboratory-derived transforms also sometimes appear in large, tectonic earthquakes (Figs. 2, 3, 5, and 6).

For each case study, we applied the transformation using fixed parameters ($\delta f = 0.1$ Hz; $\Delta f = 0.1$–0.2 Hz averaged; see *Materials and Methods* for details), followed by a two-stage smoothing filter: a backward-looking moving median (1 hour) and a moving mean (2 hours). For the 2011 Tōhoku and 2010 Maule events, these filters were broadened (4-hour median, 12-hour mean) to improve visual clarity. All transformation windows were aligned such that earthquake origin times marked the beginning of the two-minute transform interval, ensuring that the mainshock signal was excluded from the calculation. Importantly, all preprocessing and transformation steps were implemented in a strictly causal manner—using only past data within each window—to avoid any artifacts or information leakage from future samples.

As shown in Fig. 5, all five tectonic earthquakes analyzed—Gorkha (2015), Pawnee (2016), Denali (2002), Turkey–Syria (2023), and Kīlauea (2018)—exhibited distinct departures from baseline transform behavior days before rupture. These anomalies were consistently observed across multiple stations separated by tens of kilometers, highlighting their robustness.

Among these events, Denali (2002) was recorded by only one nearby station. This station was located closer to the large foreshock than to the mainshock, and its transform predominantly captured precursory activity associated with the foreshock. Notably, the signal recovered after the foreshock and again rose in the days leading up to the mainshock.

In the case of the 2023 Turkey–Syria earthquake, transform traces from all four stations analyzed showed clear pre-event excursions toward extremum values. Three of these are displayed in Fig. 5. Additionally, the transform for this event exhibited a strong diurnal modulation, suggestive of a daily episodic slow-slip-like forcing leading up to failure.

Figures 6 and S2 extends our potentially precursor analysis to two of the largest megathrust ruptures on record. In the 2010 Maule (Mw 8.8) case, a single high-quality broadband station (PLCA) registers a "wake-up" perturbation from a long-term baseline beginning nearly four months before the mainshock, followed by a rapid rise to transform extrema immediately prior to rupture. For Tōhoku (Mw 9.0), six stations (Fig. 6 panel D and Fig. S2) exhibit the same "wake-up" signal but with a different pattern: a gradual drift from long-term baselines months in advance and a pronounced run-up in the final week before failure.

In case of Tōhoku event, the dense Hi-net coverage provided a unique opportunity for field-scale validation: of the 14 nearby Hi-net sites analyzed, roughly half showed clear precursory deviations; the five strongest examples are plotted in Fig. S2. These signals remain robust despite site-specific noise and other interferences, and transforms in most frequency bands are unaffected by Tōhoku's foreshocks—acceleration consistently terminates at the mainshock.

Figure 6 also presents the transforms for the 2019 Mw 8.0 Peru earthquake, computed from more than 27 months of data at the CZSB station of the Brazil Seismic Network. In this case, the precursory patterns are among the clearest observed, appearing consistently across multiple frequency bands. The transforms reveal a sudden acceleration toward minimum values less than two weeks before the mainshock. Interestingly, another set of accelerations toward extrema—this time toward maximum values—coincides with the Mw 5.8 earthquake that occurred in the region on August 13, 2017 (Fig. 6, panels E and F).

In combination with Fig. 5, these examples confirm that the anomalous transform behavior observed in laboratory experiments prior to lab earthquakes also manifests on Earth's largest faults before natural earthquakes. Despite subtle differences between events, all examples exhibit their characteristic precursor signatures, demonstrating that our stress-sensitive transform can detect the nucleation phases predicted by laboratory experiments and simulations.

Here we emphasize that the primary aim of this study is to demonstrate that potential precursory information is present—and often visually discernible—in ambient seismic noise prior to earthquakes**.** While the proposed SSf transform sometimes reaches extrema at certain frequencies before failure, the detailed shape of SSf varies across events and settings. Accordingly, we do not claim to have identified a single, definitive precursory signal suitable for operational prediction. Instead, we view this work as a foundation for systematic follow-up: leveraging second-order statistics and cross-event comparisons to identify stable patterns, and jointly interrogating families of SSf curves to assess their consistency and transferability.

## Discussion

The novel transformation introduced in this study reveals consistent and interpretable spectral signatures of evolving fault stress. Across both laboratory and tectonic environments, we find that the transformation isolates stress-related changes in ambient seismic data, offering a promising new path toward short-term earthquake forecasting.

Crucially, the transformation is robust to amplitude variability—an advantage for analyzing ambient noise, which is often dominated by uncontrolled or distant sources. Unlike conventional velocity monitoring methods that rely on stacking or active sources, this approach provides high temporal resolution without the need for long averaging windows. This property enables near-real-time tracking of stress-induced changes in the medium, even in the absence of repeating earthquakes or triggered signals.

The observed signatures likely reflect evolving path effects as fault stress alters the elastic and scattering properties of the crust. While some contribution from source effects near the fault cannot be ruled out, the spectral slope changes captured—particularly at fine frequency resolution (0.1 Hz)—are unlikely to arise from source variations alone. Thus, the transform isolates a signal domain previously hidden in the seismic wavefield.

In several earthquakes (e.g., Turkey–Syria 2023), we observe diurnal modulation of the transform, suggesting quasi-periodic slow-slip or tidal loading effects. These clock-like patterns may offer predictive power if recurring extrema can be linked to failure probability within the 24-hour cycle. Identifying these modulations across regions could refine short-term hazard models by highlighting windows of elevated seismic risk.

These findings support a more dynamic view of earthquake nucleation. Rather than failure occurring only when a static stress threshold is exceeded—as in the classical elastic rebound model—our results suggest a process shaped by continuous stress evolution, frictional weakening, and transient loading from environmental or aseismic forcing. The consistent appearance of arc-shaped precursors across regions and magnitudes points toward a universal signature of fault readiness embedded in ambient seismic noise.

Despite these advances, not all stations analyzed in this study exhibit clear precursory signals, highlighting challenges such as:

- **Noise contamination:** Anthropogenic sources or narrow-band interference can distort the transform. Cross-station comparisons—particularly at large separations—can help confirm whether features are of tectonic origin. False positives can also be reduced through time-frequency shape analysis and multi-band consistency checks.
- **Fault complexity:** In structurally heterogeneous systems, absolute transform amplitudes may be less informative, as the transforms might capture stress fields from different faults. This requires further development of modeling frameworks that monitor the stress field on each specific fault using multiple stations.

- **Site effects:** Near-surface variability—such as nonlinear soil response or seasonal temperature shifts—can mask deeper fault processes. Borehole deployments or stations on exposed rock are less vulnerable to such distortions. Multi-station networks further mitigate local effects by enabling spatial consistency tests.

Together, these considerations support a framework in which the proposed transformation serves as a stress-sensitive filter for seismic wavefields. By monitoring subtle changes in frequency-dependent attenuation and spectral slope, the method provides new observational access to fault dynamics—potentially enabling near-real-time earthquake forecasting systems grounded in physics and validated by empirical precursors.

## Methods

*Data Preprocessing and Transform Analysis*

For the laboratory acoustic emission (AE) dataset, the data were downsampled by a factor of 4 and segmented into short, non-overlapping windows of approximately 0.016 seconds. For each segment, we computed the power spectral density (PSD) using Welch's method with a Hamming window of 2048 samples and with half window overlap, implemented via the scipy.signal.welch function. The PSD was estimated over a frequency range of 200 Hz to 500 kHz and divided into 100 linearly spaced bins, yielding a frequency resolution ($\delta f$) of approximately 5 kHz. The average spectral power within each bin was used to quantify narrowband energy content.

Using Equation 1, we computed the frequency-domain transform $SS_F(f,t)$. In cases where the frequency resolution is sufficiently fine and transforms from adjacent frequency bins exhibit similar behavior, we calculated a "local stack" by aggregating these neighboring transforms. Specifically, for local stack computations, the transform was evaluated using a bandwidth of $\Delta f = 10\delta f$. For individual transform plots shown in the Figures 1 and 2, we computed $SS_F$ using two bandwidths—$\Delta f = 10\delta f$ and $\Delta f = 11\delta f$—and averaged the results to enhance robustness. Lastly, a backward-looking moving average with a three-point window (corresponding to 0.048 seconds) was applied to smooth the resulting time series.

For the seismic datasets, only the amplitude sensitivity of each signal was corrected, avoiding full instrument response removal to prevent potential complications. Preliminary tests showed that removing the instrument response produced negligible differences in the transform results, justifying this simplified preprocessing step. The continuous waveform data were then segmented into two-minute, non-overlapping windows, and the PSD was computed using the same Welch-based method described above.

For seismic data sampled at 20 Hz, the PSD was computed over a range of 0.1–9.9 Hz using 98 linearly spaced bins, giving $\delta f = 0.1$ Hz. For data sampled above 40 Hz, the range was set to 0.2–15.2 Hz with 150 bins, also resulting in $\delta f = 0.1$ Hz. The transform was then computed for each frequency bin using $\Delta f = 0.1$ Hz and $\Delta f = 0.2$ Hz, and the two results were averaged.

To reduce noise and highlight systematic changes, we applied a two-stage smoothing procedure: a backward-looking moving median over 30 samples (equivalent to one hour), followed by a backward-looking moving average over 60 samples (two hours) for Fig. 5. To avoid contamination from post-event energy, the timing of each earthquake was aligned to the start of the corresponding two-minute window, ensuring that no post-seismic signals influenced the pre-event analysis. For Fig. 6 a backward-looking moving median over 120 samples (equivalent to four hour), followed by a backward-looking moving average over 360 samples (12 hours) is used. For slow slip in Cascadia in Fig. 2B, a 24-hour moving average is used.

Overall, the procedure used to generate the SS_f traces shown in Figs. 5 and 6 is summarized in the flowchart of Fig. 7.

**Acknowledgments.**
Seismic waveform data were obtained from the IRIS Data Management Center (DMC), the Kandilli Observatory (KO) network in Turkey, and the Hi-net network in Japan. Laboratory acoustic emission data were provided by the Penn State Rock Mechanics Laboratory (Professor Chris Marone's online repository). Data access, preprocessing, and transform computations were performed using the ObsPy Python toolbox. We acknowledge the use of PyGMT (built on Generic Mapping Tools) for creating maps. We thank Professor Gareth Funning (UC Riverside), Professor Tom Heaton (Caltech), and Zachary Zimmerman (Google) for valuable discussions that improved the manuscript. We are grateful to Professor Peter Shearer (UC San Diego), Professor Greg Beroza (Stanford University), Professor Roland Bürgmann (UC Berkeley), and Professor Chris Marone (Penn State) for their feedback on the manuscript. I also thank Professor Eamonn Keogh (UC Riverside) for encouraging me to teach an AI course, an experience that strengthened the skills later applied in developing this method. Finally, we acknowledge the use of ChatGPT (OpenAI) for assistance with code cleanup, writing support, and editorial suggestions.

**Competing interests.**
**U.S. Patent Pending**: Application No. 63/870,020, *"Methods and Systems for Detecting Earthquake Precursors via Stress-Sensitive Transformations of Seismic Noise"*, filed on August 25, 2025. The patent was filed by and is owned by the author, Nader Shakibay Senobari.

**Data and materials availability.**
All seismic and laboratory datasets are publicly available from the following sources: IRIS DMC (www.iris.edu), Hi-net (hinetwww11.bosai.go.jp), the KO network (kobezoyo.boun.edu.tr), and the Penn State Rock Mechanics Laboratory (Chris Marone's repository). All code used in this study—including ObsPy preprocessing scripts and transform routines—Code available at: https://github.com/Naderss/earthquake-precursors-ssf (archived at Zenodo, DOI: 10.5281/zenodo.16996204).

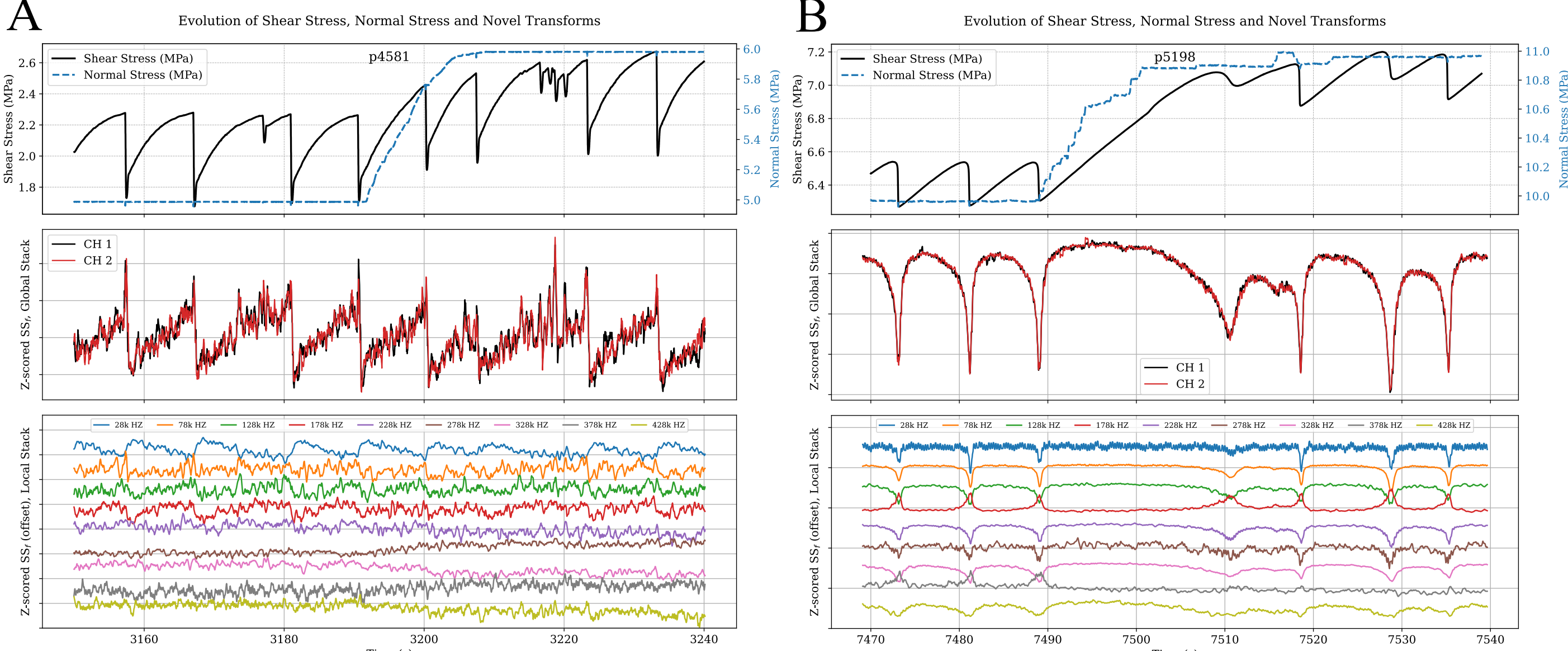

**Figure 1 | Capturing stress evolution signatures at laboratory scales using passive ultrasonic noise.** (A) *Fast-slip (stick–slip) experiment P4581*, conducted at the Penn State Rock Mechanics Laboratory, showing shear and normal stress evolution during failure of Westerly granite under triaxial loading. The top panel shows the measured stress history; the middle panel plots the stress-sensitive transform summed over all 5000 kHz frequency bins ($\Delta f = 50$ kHz); and the bottom panel displays local sums over ten-bin windows.

(B) *Slow-slip experiment P5198*, performed under lower strain rate and confining pressure, exhibits a more gradual stress release and recovery. In both cases, the passive transform captures the evolution of stress-sensitive ultrasonic energy, with distinct spectral trajectories associated with shear and normal stress variations. Together, these experiments demonstrate that passive monitoring of ultrasonic noise provides a broadband proxy for tracking stress accumulation and release—resolving both fast and slow modes of failure within laboratory materials.

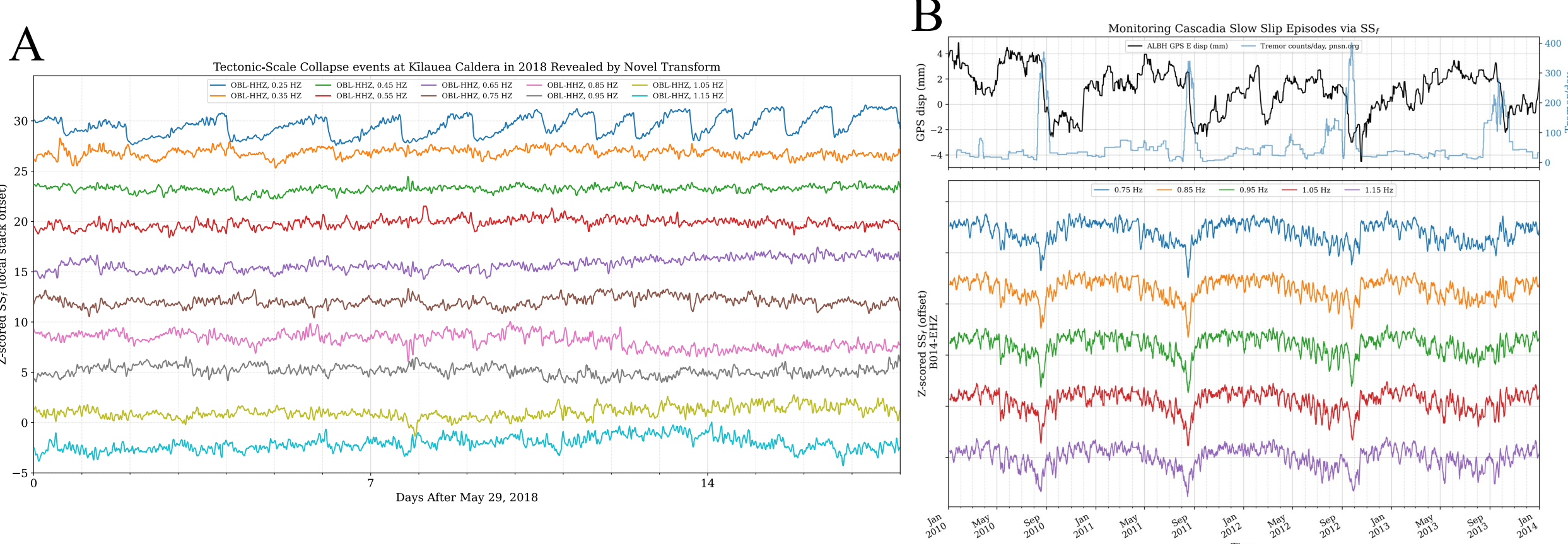

**Figure 2 | Capturing stress evolution signatures at tectonic scales using passive seismic noise.** (A) Application to the 2018 Kīlauea caldera collapse sequence using continuous seismic data from the vertical component at station OBL. The passive transform, computed in 0.1 Hz bins ($\Delta f = 0.6$ Hz) and summed locally over ten-bin windows, reveals systematic temporal evolution accompanying the collapse events. The low-frequency components exhibit an almost linear rise between successive collapses**,** followed by a sudden drop at failure, closely mirroring the stress evolution seen in the laboratory fast-slip experiment (Fig. 1A)**.** At higher frequencies, after several collapse events, a progressive shift in the baseline emerges—likely reflecting changes in normal stress within the fault zone. Moreover, the inter-event period shortens from about two days to roughly one day during this time, providing additional evidence that at higher frequencies, these transforms are sensitive to normal-stress variations.

(B) Cascadia slow-slip episodes between 2010 and 2013. The top panel shows detrended east-component GPS displacement from station ALBH (USGS) together with daily tremor counts from PNSN during Episodic Tremor and Slip (ETS) sequences beneath southern Vancouver Island. The lower panels display results from continuous records at borehole station B014 for frequency centers 0.75–1.15 Hz. During the inter-ETS period, the transform evolves smoothly along a curved trajectory resembling the upper half of a half-lemniscate, reflecting gradual changes in seismic-noise propagation associated with fault-zone loading and unloading. At the onset of ETS, it exhibits a rapid drop followed by accelerated recovery—consistent with stress-controlled modulation of seismic noise observed in the laboratory (Fig. 1). The correspondence between laboratory and tectonic observations demonstrates that the same stress-sensitive mechanism governs deformation across scales, enabling passive tracking of fault-stress evolution throughout the slow-slip cycle.

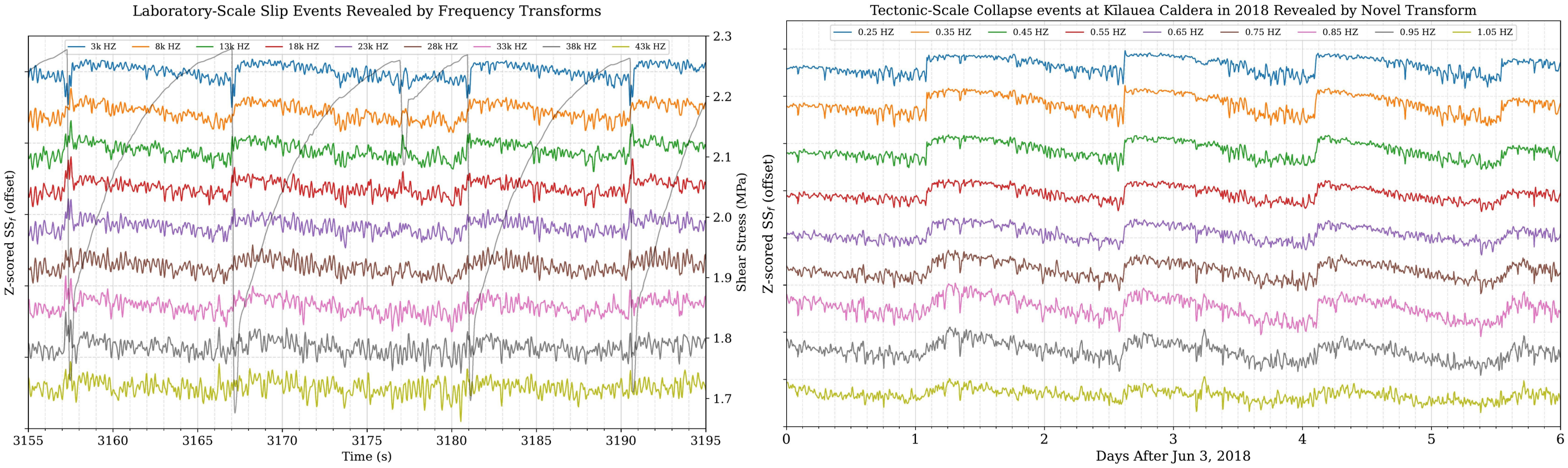

**Figure 3 | Scale-invariant shear stress signatures revealed by unstacked passive transforms.** Left, laboratory experiment P4581 (Penn State Rock Mechanics Lab) showing individual, unstacked frequency bins of the passive transform, where lower-frequency components are most sensitive to shear stress. Right, zoomed view of the 2018 Kīlauea caldera collapse seismic data (from Fig. 2A) plotted over the same frequency band. Despite a $\sim 10^{10}$ difference in event magnitude (from laboratory acoustic emissions of $M \approx -5$ to tectonic events of $M \approx 5$), the transform patterns are remarkably similar. During loading and unloading, neighboring frequencies exhibit distinct "distances to failure," ranging from near-linear to curved, arc-like trajectories. In higher-frequency bands, minima do not align immediately before failure and maxima do not follow directly afterward, producing more complex, nonlinear temporal evolution. These detailed transform behaviors illustrate how passive spectral signatures encode stress evolution, providing a physical basis for why data-driven and machine-learning approaches can successfully predict time to failure.

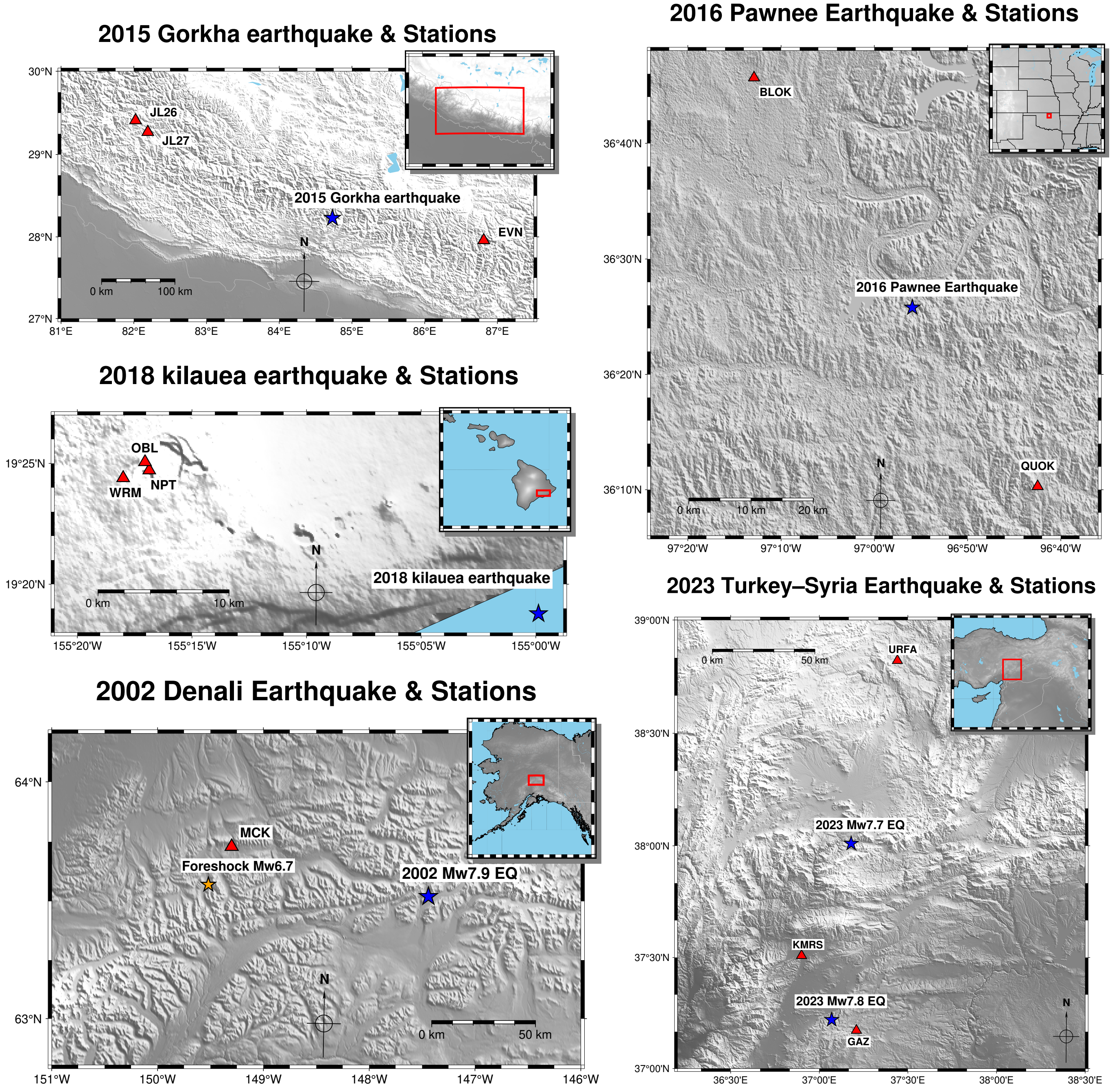

**Figure 4 | Geographic distribution of earthquake epicenters and seismic stations used in this study.** Stars mark the epicenters of the 2015 Mw 7.8 Gorkha, Nepal; 2016 Mw 5.8 Pawnee, Oklahoma; 2002 Mw 7.9 Denali; 2023 Mw 7.8 Turkey–Syria; and 2018 Mw 6.9 Kīlauea Earthquake. Triangles denote nearby three-component broadband seismic stations selected for high-quality continuous recordings. These case studies span subduction zones, intraplate faults, transform systems, and volcanic environments—demonstrating the transform's broad applicability for real-time fault monitoring.

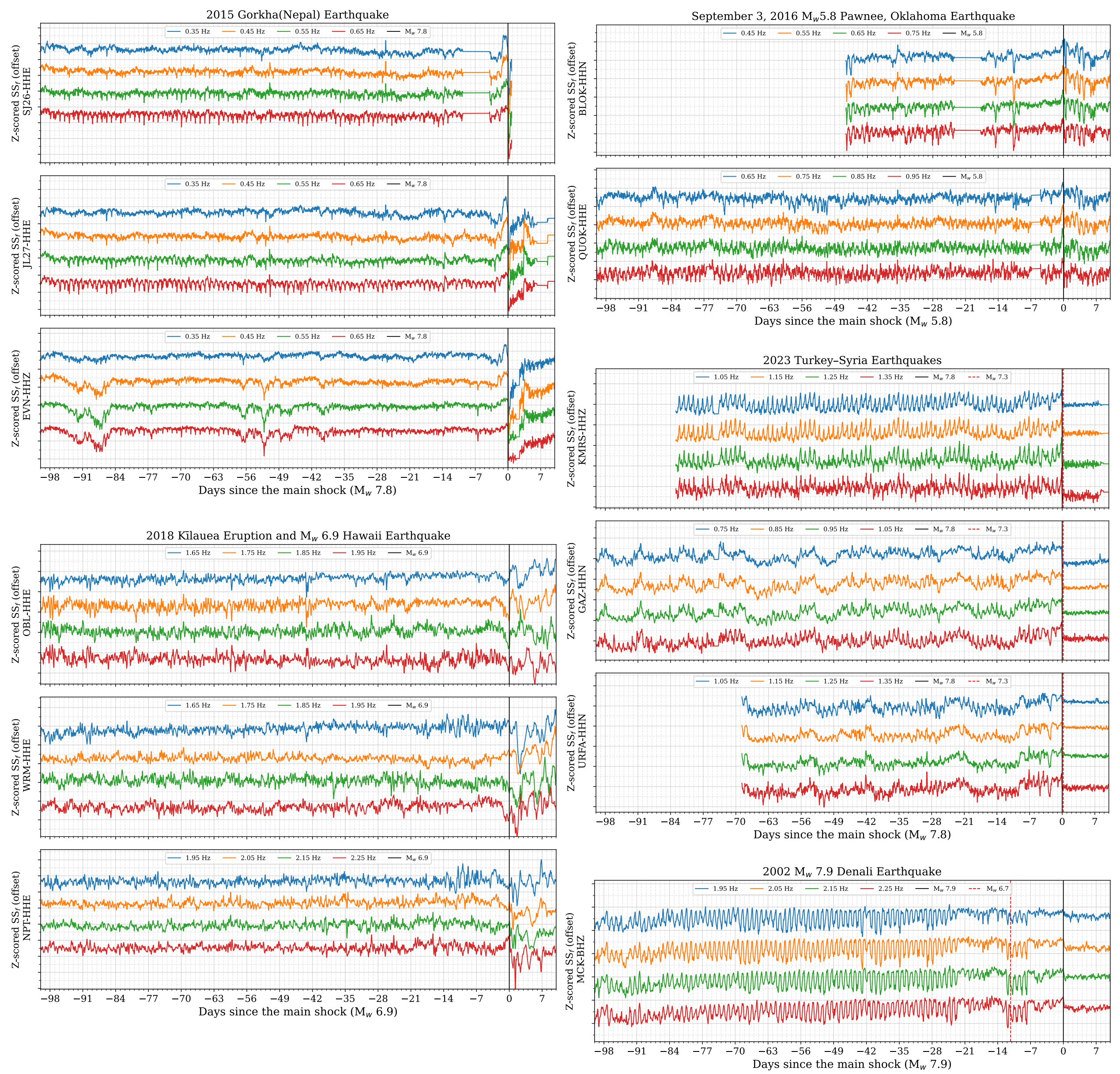

**Figure 5 | Potential earthquake precursors revealed by the proposed transformation.** Frequency-domain transforms of continuous seismic data are presented for five major earthquakes, as introduced in Figure 4. For all cases, the two hyperparameters in Equation 1 are held constant to ensure generality and demonstrate the universality of the proposed method (see *Materials and methods* for details). Earthquake origin times are indicated at the beginning of the two-minute window that includes the mainshock, ensuring that all transform values preceding this mark reflect uncontaminated pre-event signals. In all five cases, visually distinct precursor anomalies are evident, with many transform traces reaching local extrema—and, in some frequency bands, global extrema—prior to the mainshock. For clarity, only a single component of the three- component seismic data is shown: the one that displays the most visually prominent precursory features.

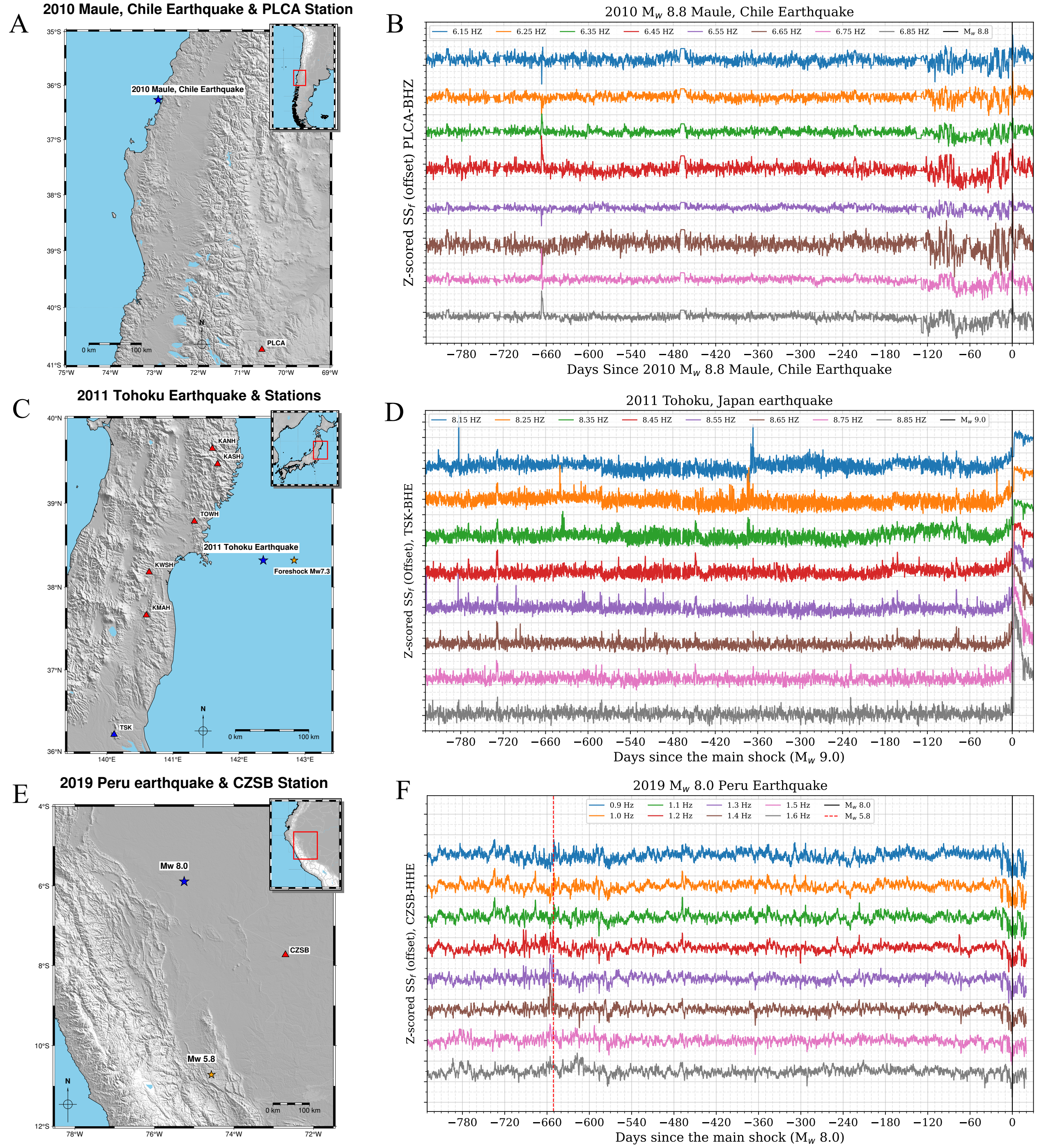

**Figure 6 | Potential precursory stress signatures preceding three large subduction-zone earthquakes. (A)** Map of the 2010 Mw 8.8 Maule epicenter (star) and the single high-quality broadband PLCA station (triangle). **(B)** Frequency-domain transform at PLCA for the Maule event, computed with the same bin width and $\Delta f$ as in Figs. 3–4 and smoothed with a 4 hr backward-looking median followed by a 12 hr backward-looking mean. The subduction zone "woke up" ~4 months before rupture, exhibiting repeated up-and-down deviations (suggesting intermittent slow-slip episodes[37]), and several frequency bands reach global extrema immediately prior to the mainshock. **(C)** Map of the Tōhoku epicenter (star), the largest foreshock (Mw 7.3; orange star), the five Hi-net stations (red triangles, transforms shown in Fig. S2), and the TSK station (blue triangle). **(D)** Transform traces at the TSK station, plotted up to 830 days before the Tōhoku event (ending at the time of instrument changes), revealing a pronounced acceleration

toward extrema values several weeks before rupture. These two case studies—among the largest recorded megathrust earthquakes—show that our stress-sensitive transform detects a sharp "wake-up" phase rather than mere gradual loading. **(E)** Map of the 2019 Mw 8.0 Peru epicenter (blue star) and the nearby high-quality broadband CZSB station (triangle). **(F)** Transform traces for the 2019 Peru earthquake, an intermediate-depth (~120 km) intraslab event, from the CZSB station. The series shows clear deviations from baseline only days before rupture, marked by a sudden downward acceleration. At some frequencies, opposite (upward) accelerations appear ~650 days before the mainshock, coinciding with a Mw 5.8 regional earthquake (orange star in panel E).

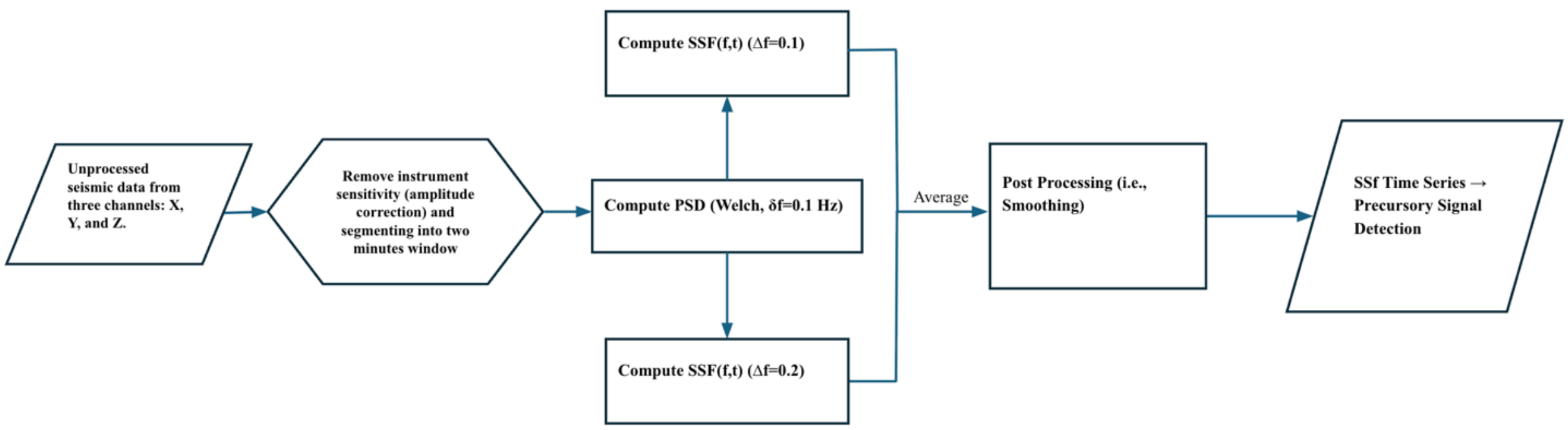

**Figure 7 | Workflow for computing SSf.** Flowchart summarizing the procedure used to generate SSf traces shown in Figs. 4 and 5. Continuous seismic waveforms are preprocessed, windowed, and transformed into amplitude spectra using Welch's method. Frequency-domain differences between adjacent bins yield the stress-sensitive transform (SSf), which is smoothed and normalized to highlight precursory features before failure.

## Supplementary Materials

Fig. S1

Fig. S2

# Supplementary Materials for

## Remotely sensing stress evolution in solids: a passive approach to earthquake monitoring

**Nader Shakibay Senobari**

Corresponding author: nshak006@ucr.edu

**The PDF file includes:**

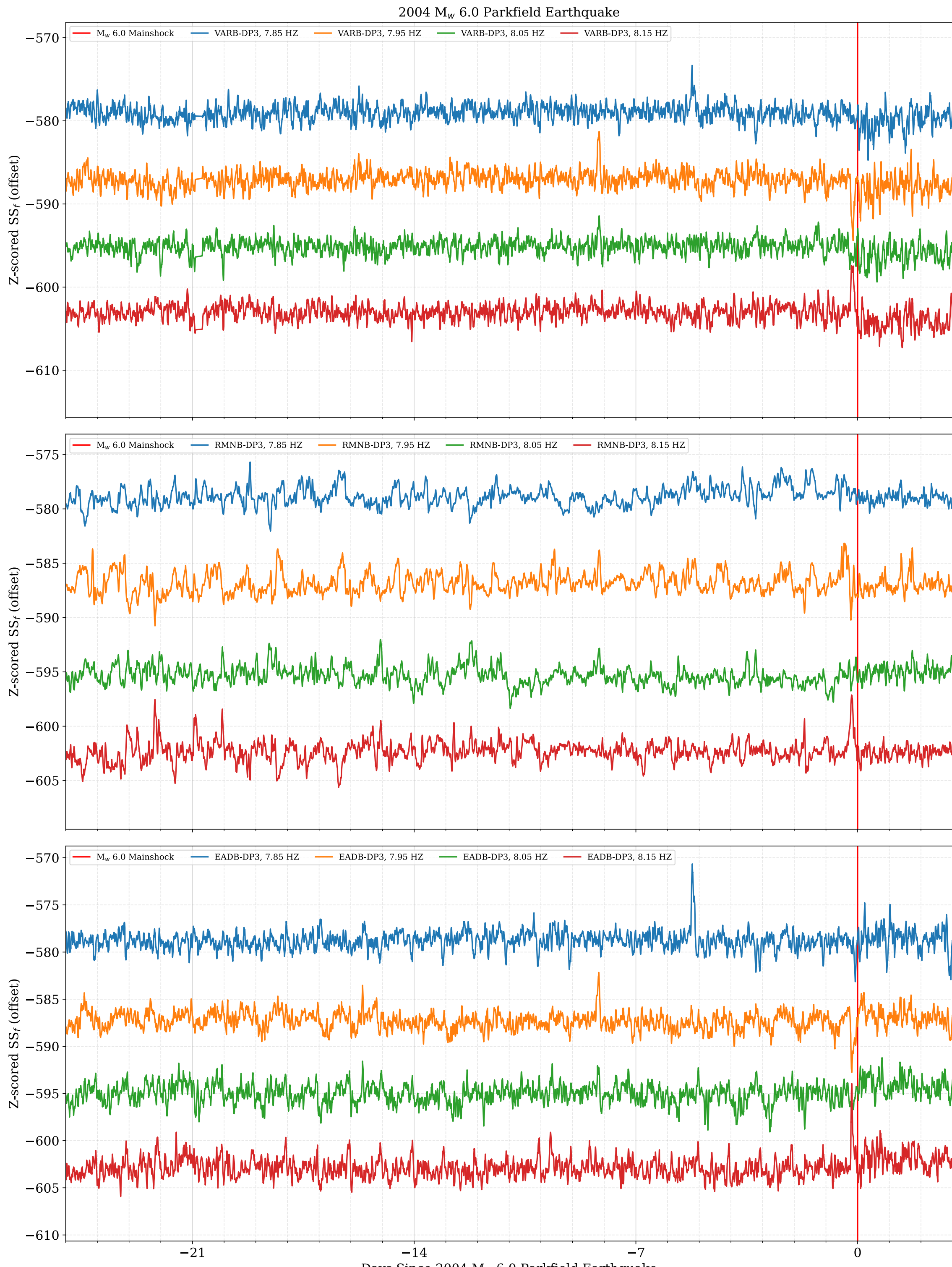

**Fig. S1.**

Novel transforms for three HRSN borehole stations in the lead-up to the 2004 Mw 6.4 Parkfield earthquake. Distinct precursor spikes appear just before the mainshock and are coherent across all three stations, indicating a multistationary signal. Similar spikes at other times lack cross-station consistency and likely reflect local site effects or instrumentation artifacts. Further analysis is required to confirm that the multistationary pre-event spikes represent genuine precursory features.

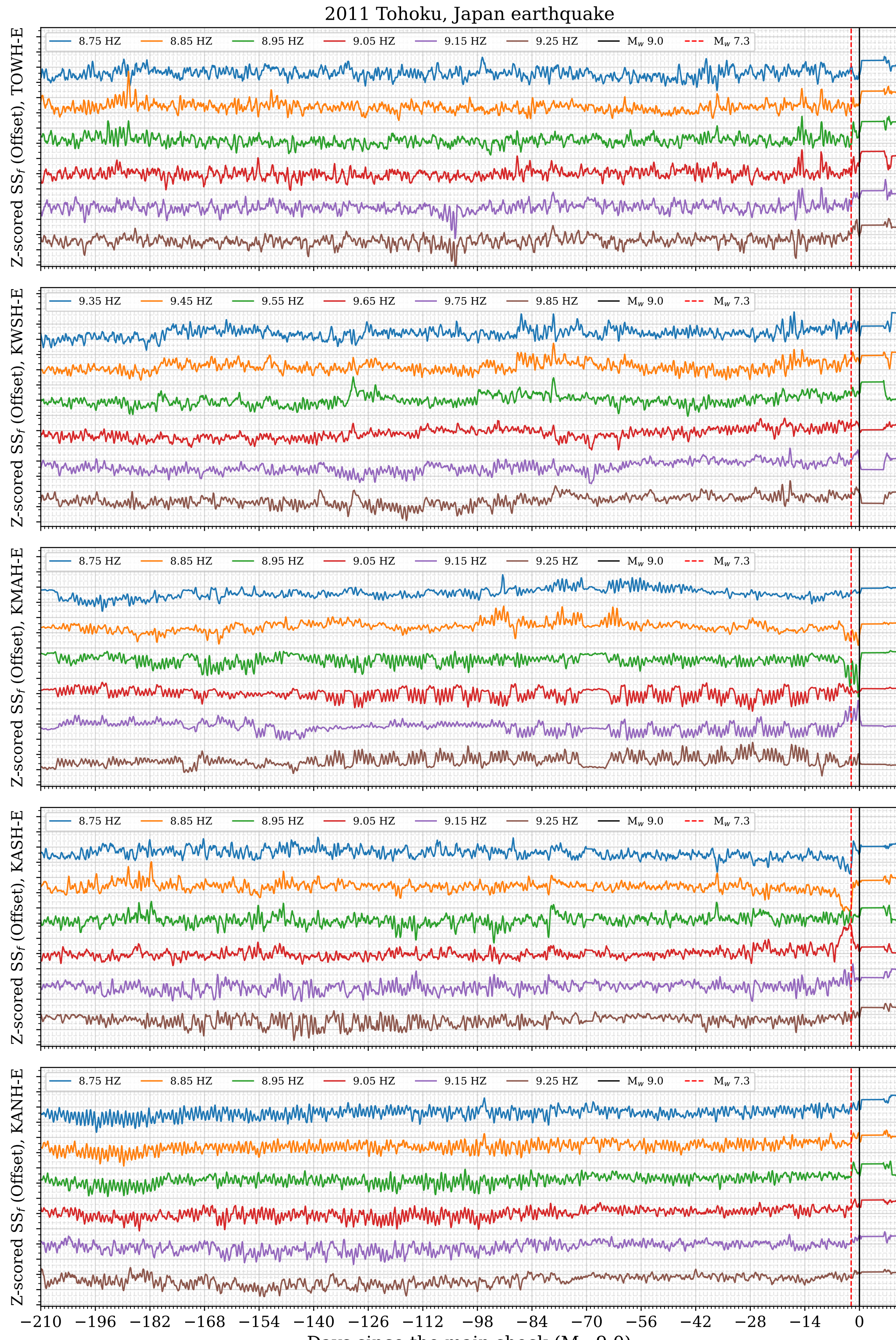

**Fig. S2.**
Transform traces for the 2011 Mw 9.0 Tōhoku event at five Hi-net stations (KMAH, TOWH, KANH, KASH, KWSH). Most series show weeks-long departures from baseline and accelerate to global extrema just before rupture; station KMAH displays the clearest precursory signal. These results demonstrate high precursor detectability across multiple stations despite local effects (nonlinear site response, nearby fault stress fluctuations, nonstationary noise, etc.).